\documentclass[conference]{IEEEtran}

\usepackage[T1]{fontenc}
\usepackage[english]{babel}
\usepackage{microtype}

\usepackage{amsmath, amssymb}
\usepackage{graphicx}
\usepackage{tabularx}
\usepackage{tikz}
\usetikzlibrary{positioning}

\usepackage{csquotes}
\usepackage{url}
\usepackage{hyphenat}
\usepackage[autolanguage]{numprint}
\usepackage[
  style=ieee,
  doi=false,
  url=false,
  isbn=false,
  eprint=false,
  maxbibnames=3,      
  minbibnames=1,      
  giveninits=true,    
  sorting=none        
]{biblatex}
\AtEveryBibitem{%
  \clearlist{publisher}%
  \clearlist{location}%
  \clearfield{language}%
  \clearfield{month}%
  \clearfield{note}%
}
\title{Trajectory-Based Recommender Systems\\ as Control Systems}

\author{
\IEEEauthorblockN{Eriam Schaffter\IEEEauthorrefmark{1},
Ahmed Bounekkar\IEEEauthorrefmark{1},
Elsa Negre\IEEEauthorrefmark{2}}
\IEEEauthorblockA{\IEEEauthorrefmark{1}Université Claude Bernard Lyon 1, ERIC, 69100 Villeurbanne, France \\
\IEEEauthorrefmark{2}Université Paris-Dauphine, PSL Research University, CNRS UMR 7243 LAMSADE, 75016 Paris, France \\
Email: eriam.schaffter@etu.univ-lyon1.fr, ahmed.bounekkar@univ-lyon1.fr, elsa.negre@dauphine.fr }
}

\begin{document}

\maketitle

\begin{abstract}
Recommender Systems (RS) are a key research domain and play an increasing role in our content-overwhelmed lives. In this paper, we explore Trajectory-Based Recommender Systems (TBRS), a subfield for which many related studies exist, yet still lacking a common framework. We argue that Control Theory provides an appropriate foundation for formalizing and solving TBRS problems. TBRS, sometimes named Long Term goal Recommender Systems, share core principles with classical RS, but at their core lies the concept of a trajectory, a defining element that makes these systems a singular category. To date, most RSs that include a notion of goal or long-term objective, when this goal is explicit, have not been recognized as having specific characteristics that make them worth regrouping under a dedicated field of research. We review related work, observe how they differ from already conceptualized RSs, and sketch the foundations of a possible theoretical framework based on control theory. Finally, we show how Educational Recommender Systems (ERS), intrinsically long-term and goal-driven, can be modeled within the proposed TBRS framework.
\end{abstract}


\section{Introduction}

A recommender system (RS) is a system that selects items from a collection and suggests them to users. RSs are a major field of research and play an increasingly important role in our content-overwhelmed lives \parencite{ricci_recommender_2022,aggarwal_recommender_2016}. They are applied in diverse domains, ranging from content recommendation on media platforms to product suggestion on commercial websites, and even people matching on dating applications. While the current state of the art identifies several distinctive categories of RSs, we argue that a significant aspect has so far been largely overlooked. It seems there is an elephant in the room of RSs.

Indeed, the literature identifies two main approaches, along with many variations, for building RSs. The two main approaches are collaborative filtering–based recommender systems (CFBRSs), based on social interactions and items evaluation by users, and content-based recommender systems (CBRSs), based on feature analysis of items and/or users \parencite{adomavicius_towards_2005,hidasi_session-based_2016}. In these methods, the RS is constructed either from the social and interactional perspective or from the feature-based perspective (content-based, using item and/or user attributes). Among the variations we find context-aware RSs, location-based RSs, knowledge-based RSs, time-sensitive RSs, and many others \parencite{negre_recommender_2017,anwar_recommendation_2025,tran_recommender_2021}. While the literature sometimes distinguishes utility-based and knowledge-based recommenders as separate categories, they can be seen as special cases or reformulations of more general paradigms. Utility-based systems rely on an explicit utility function that estimates the usefulness of an item for a user. However, such a function is implicit in virtually all recommender systems, since any predicted rating or ranking corresponds to an estimation of utility. Similarly, knowledge-based systems encode explicit domain knowledge about items and/or user needs, but their reasoning mechanisms often overlap with content-based approaches. Therefore, both can be viewed as specific instantiations or extensions of the broader content-based or collaborative filtering paradigms, rather than as fundamentally distinct categories. Also, Context-Aware RSs usually leverage additional information such as location, season or social data to improve recommendations. This additional information mainly concerns the user and/or the items and can be seen as an evolutionary step in recommender systems of that time. Historically, it is worth noting that research on recommender systems initially focused on short-term, engagement-oriented recommendations before gradually expanding to encompass other objectives \parencite{bogina_considering_2023}.

But there is also an additional category of recommender systems specifically designed to guide users toward a defined goal, for example Educational Recommender System (ERS) or Point-of-Interest Recommender System (POIRS) used in tourism: we except that these systems will ``recommend a sequence'' \parencite{ricci_recommender_2022}. While related to existing approaches, it differs in a fundamental way \parencite{hossein_nabizadeh_long_2015,zhang_learning_2022}. We argue that, in this category of RSs, dynamics, and more specifically trajectories, are the central constituent of the RS. As occasionally found in the literature, such systems are referred to as Trajectory-Based Recommender Systems (TBRSs) or Long-Term Goal Recommender Systems (LTRSs). In this paper, we adopt the TBRS denomination to emphasize that the core of the system is the trajectory thereby making this denomination appropriate for a whole class of problem resolutions and, in our view, the term trajectory is preferable to path, as it conveys a stronger association with dynamic processes, whereas path suggests a more static notion \parencite{wiktionaryorg_trajectory_2024, wiktionaryorg_path_2025}. Yet, we will occasionally use the term path to remain consistent with the terminology employed in the cited papers. Several existing systems could already be interpreted as TBRSs \parencite{hutchison_trajectory-based_2012,huang_using_2012,tran_recommender_2021,barbaric_health_2025} in tourism, education or even health. And while it seems that numbers of RSs inherently contain a trajectory dimension and that RSs can produce trajectorial effects on users, relatively to the choosen utility function of the RS, neither the trajectorial effect nor the trajectory dimension only makes a TBRS.  
In our view, a TBRS is a RS in which the trajectory is at the center of the system, not only a dimension, and we argue that framing this problem within the formalism of Control Theory (CT) enables both the application of established solutions and the evaluation of outcomes of these systems. The central research question is whether CT can effectively be appropriate for modeling TBRSs.

The remainder of this paper is organized as follows. Section II reviews related work, identifies the emergence of trajectory-based recommender systems in various domains and formalizes the concept of TBRSs. Section III introduces a control-theoretic formulation providing a unified mathematical framework for TBRSs. Section IV illustrates its application to ERSs. Finally, Section VI concludes and discusses future research directions.

\section{Related Work}

Since the first mention of this model by \textcite{goldberg_using_1992}, originally intended to facilitate the reading of electronic mail, up to the Netflix Prize, the history of recommender systems (RS) has been marked by rapid advances aimed at bringing adequate content directly to users, before they explicitly search for it \parencite{resnick_recommender_1997}. The field of RSs remains highly active, with applications ranging from content-streaming platforms and social networks to education \parencite{tadlaoui_educational_2018}, health education \parencite{vandeputte_coaching_2022}, and tourism \parencite{picot-clemente_generic_2011}. 

In essence, a RS consists of a set of resources (the \textit{items}) 
\( R = \{ r_1, r_2, \dots, r_m \} \) with  \( m \in \mathbb{N^+} \), a recommendation engine (RE), a user interface (UI) and a \textit{user} and/or a set of \textit{users}. 

In the literature, creating a recommendation is suggesting within \( R \) a subset \( R^* \subseteq R \) containing the item the \textit{user} will most likely make a decision with, that is maximizing a relevance function \( f(user, r_i) \): 
\[
R^* = \operatorname*{arg\,max}_{r_i \in R} f(user, r_i),
\]
where \( f(user, r_i) \) represents the relevance or utility of item \( r_i \) for \textit{user}, with \( i \in \mathbb{N} \) and resources defined in dimension \( m \) \parencite{ricci_recommender_2022}. The decision usually involve choosing a music to listen, choosing a news to read or a product to buy but it can also be which point of interest to reach or which food to order while the relevance is usually related to the on-going engagement of the user in the RS. Various techniques coexist to provide recommendations to users.

\subsection{Classical Recommender Systems}

In the classical methods for items recommendation we first find the principle of recommendation based on Neighborhood-Based Collaborative Filtering (NBCF). It relies on \textit{ratings} provided by \textit{users} on \textit{items}. It assumes that users with similar preferences tend to rate similar items in similar ways \parencite{aggarwal_recommender_2016}, thereby allowing the prediction of ratings for items that a user has not yet evaluated.

Two main approaches of NBCF are User-Based Collaborative Filtering (UBCF) and Item-Based Collaborative Filtering (IBCF). UBCF recommends an item to a user based on the similarity of that item’s ratings among other users who share similar evaluation patterns across other items. Conversely, IBCF recommends an item to a user by considering the similarity among a set $S$ of items and the ratings the user has assigned to the items in $S$. In both methods, the predicted rating is typically computed as a weighted average of the neighbors’ ratings \parencite{aggarwal_recommender_2016}. In these approaches, both algorithms can be used either to predict a \textit{rating} for a specific \textit{user–item} pair, or to generate a \textit{top-$k$} list identifying the $k$ most relevant items or users. In practice, generating a \textit{top-$k$ items} list to recommend to a user is often more useful than predicting an individual rating value.

Second the RSs based on the \textit{features} of the items constitutes what's called Content Based Recommender Systems (CBRS). Many variations of theses systems exists and they share the basic principle of an approach that consists of matching features of the item side to features of the user side (preferences of the user). It usually uses historical and implicit or explicit preferences of the user and extrapolates, based on the features of items, to which items the user may be attracted to \parencite{ricci_recommender_2022}. These systems can often take advantage of a semantic enrichment, though the use of ontologies for example, to offer better recommendation outcomes.

Then, hybrid approaches tend to mix various techniques to produce an adequate list of recommended items and to overcome the limitations of individual methods  \parencite{ricci_recommender_2022}. Context-Aware Recommender Systems (CARS) can also provide additional information improving the relevance of recommended items based on context related to the user and/or the items \parencite{ricci_recommender_2022}.

Finally, numerous evaluation metrics assess RS performance, most of which focus on the top-$k$ recommended items, the typical output of RSs, regardless of whether they rely on collaborative, content-based, hybrid, or context-aware approaches.

\subsection{Trajectory-Based Recommender Systems}

Then, by adding to a RS a notion of goal the paper Long Term Goal Oriented Recommender Systems by Nabizadeh et al. \parencite{hossein_nabizadeh_long_2015} introduced the fundamental concept of a RS that actively drives a user towards ``a target area of the item space''. While the importance of Long-Term Recommender Systems (LTRS) is clearly articulated in this work, the opportunity to investigate the problem independently from the influence of classical RS paradigms seems to have been overlooked. 

In the same paper, among the promising applications of LTRS discussed, the authors identify educational recommendation as a particularly fertile domain. In this paper Nabizadeh et al. formulated the idea of an educational LTRS, an idea presented before as the Learning Path Recommendation (LPR) in \cite{durand_graph_2013}. More recently the research community focus widely on the long term goal in Educational Recommender Systems (ERS) and devised the recommendation task as a sequential decision process \parencite{huang_deep_2021, zhang_learning_2022}. 

In other domains, path-based recommendation has also been extensively studied \parencite{hutchison_trajectory-based_2012,liu_pathway-finder_2014,yuan_improving_2021}. Tourism, for instance, offers a natural setting in which points of interest (POIs) can be sequenced into a personalized itinerary adapted to each visitor \parencite{hutchison_trajectory-based_2012}. Health care is another domain where path recommendations are particularly relevant. The sector is characterized by both its imperative for mass customization and its commitment to quality of care. Clinical pathways, in particular, are designed to provide patients with trajectories that combine standardization for safety and efficiency with customization for individual needs \parencite{huang_using_2012}.

Recently, Bogina et al. \parencite{bogina_considering_2023} reviewed extensively the temporal aspects in RSs, they underlined the fact that the time dimension should be seen as either a container for short term preferences or a container for long term preferences of the user. The paper mention the Four-Phase Model of Interest Development that includes four sequential phases: an initial situational interest, a maintained situational interest, an initial individual interest and finally a developed individual interest which somehow defines a journey through the evolving interests of the user. But having a time-aware RS does not necessary means building a TBRS: considering time in a RS, to improve the quality of recommendations, does not imply considering a goal in the item space. Similarly considering sequences of user choices \parencite{meng_user_2025}, to improve the quality of recommendations, does not imply considering a goal in the item space. The notion of goal is essential in a TBRS and the field of requirement engineering defines a goal as ``expressions of intent and thus, declarative with a prescriptive nature'' \parencite{aurum_modeling_2005} similar to what a trajectory is made to achieve.

So, unlike general-purpose RSs, which focus on satisfying the user’s immediate preferences, we are interested in systems explicitly addressing long-term objectives. These RSs aim to optimize a process of some sort, effectively distinguishing themselves by their notion of goal from classical short sighted RSs often optimizing for engagement and therefore sometimes fall short of serving the user’s true benefit \parencite{anwar_recommendation_2025}. The type of process being optimized is not restricted as long as it involves taking decisions. While optimizing a specific process these RSs should also exhibit features according to what is academically considered being a recommendation.

First, what makes a TBRS a RS is that it constitutes a complete system, comprising a specific design, a dedicated user interface (UI), and a recommendation engine (RE) \parencite{ricci_recommender_2022}.  Within the framework presented here, it is explicitly assumed that it constitutes a whole system which goal is to guide the user along a trajectory towards a desired state. Second, a TBRS usually targets users who either lack the competences or the experience required to evaluate which items are relevant \parencite{ricci_recommender_2022}, in our case, users are expected to make a sequence of decisions while lacking prior knowledge of the most adequate decision to take at the most adequate moment. Third, classical RSs typically produce a ranked list of individual items and a TBRS also produces a ranked list of items, but these items collectively and sequentially organized constitute a trajectory. Finally a TBRS is user-centered, a single trajectory for a user is potentially different from a trajectory of a different user. Finally it also operates within a large item space to reduce information overload \parencite{ricci_recommender_2022}. The Table \ref{tab:rs_vs_tbrs} summarize these findings.

\begin{table}[h]
\centering
\renewcommand{\arraystretch}{1.2}
\begin{tabularx}{\columnwidth}{|X|c|c|}
\hline
\textbf{Feature} & \textbf{Classical RSs} & \textbf{TBRSs} \\
\hline
Complete system 
& Yes 
& Yes \\
\hline
Has a trajectory
& Implicit
& Explicit, central \\
\hline
Target novice users 
& Yes
& Yes \\
\hline
Decision process 
& One-shot 
& Sequential \\
\hline
Recommendation output 
& Ranked list 
& Ranked list \\
\hline
Handling large item space 
& Yes
& Yes \\
\hline
\end{tabularx}
\caption{Comparison between RSs and TBRSs \parencite{ricci_recommender_2022}.}
\label{tab:rs_vs_tbrs}
\end{table}

Thus, by comparing these features, we can identify the two key differences that distinguish TBRS from classical RSs: first, the central role of goal orientation in TBRS, and second, its consequential impact on the user's decision process, which becomes inherently sequential rather than one-shot. One side remark, however, is that some RSs can also be useful for informed users, in addition to being beneficial for novice users, particularly in situations where information is either sparse or overly abundant \parencite{tran_recommender_2021}.

A review of the literature reveals that these conceptual features are reflected in a wide range of existing works, as summarized in Table~\ref{tab:tbrs_features_refs}. This literature review followed an exploratory approach rather than a systematic methodology, aiming to identify representative works illustrating each conceptual feature.

\begin{table}[h]
\centering
\renewcommand{\arraystretch}{1.2}
\begin{tabularx}{\columnwidth}{|X|c|}
\hline
\textbf{Illustrated feature} & \textbf{Representative references} \\
\hline
Complete system 
& \cite{hafsa_multi-objective_2022, amir_hossein_nabizadeh_rutico_2017}  \\
\hline
Has a trajectory
& \cite{golbeck_recommender_2025, ferrero_narrative-driven_2025, du_not_2025, anwar_recommendation_2025}  \\
\hline
Target novice users 
& \cite{daniel_automated_2010} \\
\hline
Decision process 
& \cite{zhang_learning_2022, liaw_reveal_2022} \\
\hline
Recommendation output 
& \cite{henning_peter_a_learning_2018} \\
\hline
Handling large item space 
& \cite{meng_user_2025} \\
\hline
\end{tabularx}
\caption{Illustrated TBRS features found in the literature}
\label{tab:tbrs_features_refs}
\end{table}

A TBRS transcends the notion of a simple path-finding algorithm, even though a study evoke a depth-first search as an possible solution \parencite{amir_hossein_nabizadeh_rutico_2017}. While preserving the fundamental characteristics of RSs, it differentiates itself from classical RSs through the distinct behavioral expectations placed upon it: a TBRS mediates the user's need for engagement on the platform through a more subtle, goal-oriented utility function. This distinctiveness substantiates the need for a dedicated and rigorous conceptual model at the core of such systems.

\section{A Control-Theoretic Formulation}

As we just discussed, classical recommender systems (RS) are typically designed for short-term optimization, suggesting a set of items that maximize immediate user utility. 
In contrast TBRS aim to guide a user from an initial state to a desired target state, thus optimizing over a longitudinal trajectory rather than a single decision. Numerous approaches has been tested to solve this problem, amongst these we find particle swarm optimization \parencite{subiyantoro_learning_2021, xiaojing_sheng_adaptive_2023}, reinforcement learning \parencite{huang_deep_2021, zhang_learning_2022} or deep learning \parencite{zhou_personalized_2018}. To our understanding, however, existing methods lack a global, coherent, simple, and comprehensive formulation. In the authors’ view, this is partly due to a conceptual jump in considering the user in a TBRS as being \textit{guided} through sequential choices and progressively drawn toward a goal: considering the user as a dynamical system. Acknowledging the idea of purposefully guiding sequential choices and, more generally, \textit{guiding} the user, viewed as a dynamical system, toward long-term outcomes opens the way to framing the problem within a unified theoretical model.

And this formulation resonates with a well-established framework in the control of dynamical systems, whether mechanical, biological, or social, that has not yet been systematically explored for recommender systems: Control Theory (CT). CT provides a rigorous and extensible mathematical foundation, which is not restrictive on solution methods, which naturally accommodates uncertainty, and which directly aligns with the requirements of TBRS: steering a dynamical system (the evolving user state) from an initial state $x_0$ to a target state $x_{target}$, through a sequence of control inputs (what will become the recommendations in the RS), while preserving agency and choice for the user.

CT is also appropriate for a RS because it can accommodate various control strategies. Robust control is suitable when significant discrepancies exist between the model and the real system, and the controller must ensure fail-safe performance. Optimal control aims to determine the best possible control inputs according to a predefined cost function. Stochastic control deals with uncertainty and is conceptually close to Partially Observable Markov Decision Processes (POMDP). And, finally, intelligent control leverages various artificial intelligence approaches to govern a dynamical system.

Using CT and translating it to our field of problem, in its simplest linear formulation \parencite{rawlings_model_2017}, we could express a TBRS dynamics as:
\begin{equation}
    x(t+1) = A x(t) + B u(t)
    \label{eq:state_dynamics}
\end{equation}
where $x(t) \in \mathbb{R}_+^n$ denotes the user’s state at time $t$, and $u(t) \in \mathbb{R}^m$ represents the recommendations (We will later discuss how to derive the top-$k$ items from the optimization problem and how to incorporate the user’s choice into the system dynamics). The matrix $A \in \mathbb{R}^{n \times n}$ models the natural evolution of the user’s state (its interpretation may vary depending on whether the TBRS concerns education, tourism, or health), while $B \in \mathbb{R}^{n \times m}$ captures the effect of the recommended items and is constituted by the \textit{resource base}. These matrices may be heuristically designed, empirically estimated, or learned from data. 

\subsection{Solving with a Model Predictive Control}

This formulation directly enables the use of standard control-theoretic tools. Remember that in a TBRS we place a focus on the trajectory and its direction towards a goal. As we discussed one of the option for solving our TBRS dynamic is to optimize the trajectory over a cost function. We can build a TBRS upon an optimal control approach and we can either solve the entire path with a Linear–Quadratic Regulator (LQR) \parencite{rawlings_model_2017} or solve a portion of the path and iterate towards the goal with Model Predictive Control (MPC) \parencite{rawlings_model_2017}. MPC performs a partial optimization of the control sequence over a finite horizon and is well suited to accommodate constraints. It provides an optimal set of controls at time t: $U^*(t)$ consisting of the sequence of optimal controls $[u_0(t), \dots, u_T(t)]$ for each step up to the finite horizon $T$.

Therefore, using MPC constitutes an appropriate approach. We will discuss this method as an example of how to compute the optimal recommendation trajectory, which can be expressed as the minimization of a quadratic cost function over a prediction horizon \( T \):

\begin{equation}
\begin{aligned}
U^*(t)
&= \underset{\{u(\tau)\}_{t}^{t+T-1}}{\operatorname{arg\,min}}
\Bigg\{
\sum_{k=0}^{T-1}\!\big(
\mathbf{x}_{t+k}^{\top} Q\, \mathbf{x}_{t+k}
+ \mathbf{u}_{t+k}^{\top} R\, \mathbf{u}_{t+k}
\big)
\Bigg\}                                              \\
&\quad +\, \mathbf{x}_{t+T}^{\top} Q_f\, \mathbf{x}_{t+T}.
\end{aligned}
    \label{eq:optimization}
\end{equation}

where $Q \in \mathbb{R}^{n \times n}$ weights the deviation of states (encouraging fast convergence), 
$R \in \mathbb{R}^{m \times m}$ penalizes excessive or costly recommendations, 
and $Q_f \in \mathbb{R}^{n \times n}$ emphasizes the quality of the final state. 

The resolution of this optimization yields an optimal sequence of items guiding the user along a trajectory toward the target state. At each interaction, the user’s state $x(t)$ is updated, and a new optimization is solved, making the TBRS an iterative and adaptive RS grounded in CT.  Moreover, this formalism is extensible: non-linear dynamics, stochastic formulations, or hybrid models can be employed to capture richer user behaviors while constraints on controls input can satisfy real-world or heuristics limitations. Constraints can also encode optimization heuristics derived from model knowledge, effectively implementing a more efficient trajectory-based model.

\begin{figure}
    \centering
    \includegraphics[width=1\linewidth]{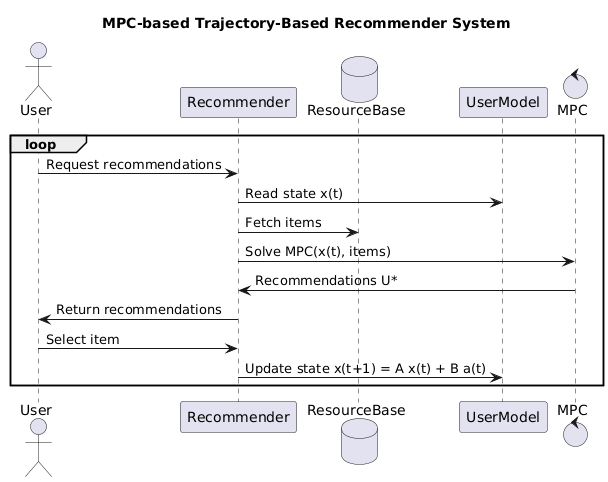}
    \caption{Sequence diagram of the TBRS}
    \label{fig:seq_diagram}
\end{figure}

But, unlike traditional CT systems, and due to the nature of our system as a RS, the human is in the loop. And while the construction of the control input vector \(u(t)\) requires the formulation of an MPC optimization problem, instead of directly applying this control vector to the system, only a single resource from the resource base could be used by the user at time $t$ as we consider that any use of the RS consists of the choice of a single resource (and in its effective use by the user). So we suggest that $U^*(t)$ will be used to generate the Top-\(k\) items in the recommender and that the user’s choice, here illustrated by \(a(t)\) will then directly influence the system dynamics through the action vector as illustrated by Figure~\ref{fig:seq_diagram}.

\subsection{Generating the top-$k$ items}

So, as each optimization occurs, at each time $t$ we have a sequence \(U^*(t) = [u_0(t), \dots, u_n(t)]\) and, as we just observed, in the control of automated dynamical systems, \(u_0(t) \) is usually applied \textit{all at once} which means that each element of \(u_0(t) \) is multiplied by $B$ to produce the desired effect on the dynamical system. Since we are building a RS we preserve user agency in the choice of the single control to apply, effectively translating one element of the generated Top-\(k\) into the single action \(a(t) \) applied to the dynamics.

This leaves us with the question of the generation of the Top-\(k\) items from \(U^*(t)\). As a simplification we can built it long-sighted or short sighted depending on any combination of the items evaluated during the optimization and existing in \(U^*\). That means we can built the Top-\(k\) only on $u_0(t)$ or on \(u_0(t) .. u_T(t)\) following a decision of offering to the user more items in the pool of the immediate controls or in the pool of subsequent controls. Selecting whether to construct the Top-\(k\) recommendations from the instantaneous control \(u_0(t)\) or from the entire optimized sequence \(U^*\) represents an interesting policy design question that may depend on multiple contextual factors. Intuitively, as in TBRS, combining short-term recommendations driven by the current state with long-term guidance derived from the optimized trajectory could lead to improved performance.

Below is an example optimization for a set of four controls over an horizon of five. We can see that we have a decision to make over the choice of controls selected to build a Top-\(k\):

\[
U^*(t) =
\begin{bmatrix}
u_{0,1}(t) & u_{1,1}(t) & u_{2,1}(t) & u_{3,1}(t) & u_{4,1}(t) \\
u_{0,2}(t) & u_{1,2}(t) & u_{2,2}(t) & u_{3,2}(t) & u_{4,2}(t) \\
u_{0,3}(t) & u_{1,3}(t) & u_{2,3}(t) & u_{3,3}(t) & u_{4,3}(t) \\
u_{0,4}(t) & u_{1,4}(t) & u_{2,4}(t) & u_{3,4}(t) & u_{4,4}(t)
\end{bmatrix}
\]

We can produce a Top-\(k\) by solving the Hadamard product\footnote{The Hadamard product is an element-wise matrix multiplication that takes two matrices of the same dimensions and multiplies each element of one matrix by the corresponding element of the other.} of our $U^*(t)$ with a one-hot projection matrix that can satisfy various policy. For example a short term policy that would create a Top-\(k\) out of the first two set of controls: 

\[
P_K^{\text{short}} =
\begin{bmatrix}
1 & 1 & 0 & 0 & 0\\
1 & 1 & 0 & 0 & 0\\
1 & 1 & 0 & 0 & 0\\
1 & 1 & 0 & 0 & 0
\end{bmatrix}
\qquad
Top_k^{\text{short}}(t) = U^*(t) \odot P_K^{\text{short}}
\]

Or another policy that would build a medium sighted Top-\(k\):

\[
P_k^{\text{med}} =
\begin{bmatrix}
1 & 0 & 0 & 0 & 0\\
1 & 1 & 0 & 0 & 0\\
1 & 1 & 1 & 0 & 0\\
1 & 1 & 1 & 1 & 0
\end{bmatrix}
\qquad
Top_K^{\text{med}}(t) \;=\; U^*(t)\ \odot\ P_k^{\text{med}}
\]

Then, provided with a Top-$k$ the user can select an \textit{item} and the action chosen by the user among the Top-\(k\) items is then applied to the user model according to:
\[
    x(t+1) = A\,x(t) + B\,\big(u(t) \odot a(t)\big),
\]
where \(a(t)\) is a one-hot vector identifying the selected resource. This formulation preserves the predictive control structure while introducing the human-in-the-loop selection at each step. Finally and iteratively the optimization process is ran again as illustrated by the sequence diagram \ref{fig:seq_diagram}. 

In essence, this perspective positions TBRS as a control problem on user trajectories, offering a principled way to design RSs that are not merely reactive but goal-directed, adaptive, and optimally sequenced. 

As with most classical RSs, a TBRS can be evaluated using standard ranking metrics such as precision, recall, Root Mean Square Error (RMSE) or the Normalized Discounted Cumulative Gain (NDCG) \parencite{michiels_best_2024}. In addition, several domain-specific evaluation approaches have been proposed, many of which are closely tied to the application context, for example measuring the distance from top learners in \cite{zhang_learning_2022}. The proposed framework for TBRS has yet to be consistently evaluated for its relevance with these common metrics on offline datasets.

\section{An application in education}

\subsection{Framing the Control-Based Learning Solution}

As Nabizadeh et al. \cite{hossein_nabizadeh_long_2015} and others have stated, education is a perfect field for building TBRSs, the learner needs agency \parencite{guay_applying_2022} and need to acquire competencies over time. To illustrate the applicability of the control-theoretic formulation to educational recommender systems (ERS), we would consider a learner interested in improving his competences.

So, if we use the system dynamics Equation \ref{eq:state_dynamics} to represent the learner’s evolution we could use $A$ to model the natural progression (or forgetting dynamics, for example we could apply a decay similar to Ebbinghaus' curve \parencite{murre_replication_2015}), we could use $x(t) \in \mathbb{R}^n$ to encode the learner's state vector and thus the estimated mastery level over $n$ competences, then in $B$ we could encodes the pedagogical effect of each resource coming from a resource base on each competence dimension, $B$ would have a size of $\mathbb{R}^{n*m}$. In $B$, each row represents a competence and each column corresponds to a specific learning resource.

Now, we want to recommend a sequence of learning items that drives measurable progress over time. So we want to produce a Top-$k$ recommendation items set based on \(U^*\) as discussed above and using a medium policy of projection. 

To generate a list of recommendations for our learners, we recall that our recommender system (RS) is trajectory-based: it predicts and optimizes the learner’s evolution over time. In order to compute this trajectory, we must specify a target state that represents the desired level of mastery for each skill. This target vector encodes the competencies that our student is expected to acquire by the end of the learning period. 

The goal of the trajectory-based recommender is thus to determine the optimal sequence of learning items that minimizes the gap between a learner’s current skill vector \(x(t)\) and the target \(x_{\text{target}}\) over the prediction horizon. We can then use an MPC optimization to build such a sequence.

In our educational setting, the MPC formulation governs the allocation of learning resources in order to guide learners toward a desired target skill set. We can use the optimization presented in Equation \ref{eq:optimization} and solve to obtain our optimal \(U^*(t)\). In this MPC formulation the target state does not play a direct role in the optimization though, it serves only as a reference. In the case where the target state should be reached, during the optimization, it should be stated as an additional constraint in Equation \ref{eq:optimization}.

With this approach, once the optimal control sequence \(U^*(t)\) has been computed, the system can generate an ordered list of recommended resources. Each component of our Top-$k$ corresponds to a whole or a part of each column of the matrix \(B\), that is, to a specific pedagogical resource or learning action. The resources are ranked according to their efficiency in driving the user’s current state \(x(t)\) toward the desired goal \(x_{\text{target}}\).

In effect, in the user interface, our learners are presented an ordered list of items. They can then select the one that best suits their preferences at time \( t \). By making such a choice, the learner effectively selects one resource, and we infer from this choice an action \(a(t)\), corresponding to the chosen action. We consider, as a simplification, that consuming a resource equals acquiring the level of competence specified by this resource 

So finally, in the user interface, our learners sees the ranked list and picks one item. After the user action has been applied and the new state $x(t{+}1)$ is obtained, the system can iterate and re-optimize the control sequence $U^*(t{+}1)$ using the updated state of the learner.

This receding-horizon process continuously adapts recommendations to the evolving competence profile as illustrated in the Figure \ref{fig:seq_diagram}.
Through this loop of prediction, action, and re-optimization, the system behaves as a dynamic TBRS, where educational trajectories are generated in real time based on the learner's feedback and choices.

\subsection{Simulating and Evaluating Various Learning Scenarios}

We test three different scenarios of our MPC-based solution using synthetic data and a number of simplifying assumptions. The first scenario assumes mono-competence resources, where each action primarily affects a single skill dimension. The second scenario introduces semi-clustered competences, allowing limited overlap between related skills. The third scenario considers multi-competence resources, where each action influences several skills simultaneously.

We chose to simulate the learner's choice as probabilistically oriented toward the highest-ranked recommendation, while introducing a small stochastic component within the Top-\(k\) set to reflect natural variability in learner behavior. We also experimented with various parameterizations of $Q$ and $Q_f$, while assigning higher weights in $R$ to resources categorized as more difficult. Finally, we generated 80 random resources to allow the solver to optimize the trajectory over a small but reasonable number of items.

\begin{figure}[h]
    \centering
    \includegraphics[width=\linewidth]{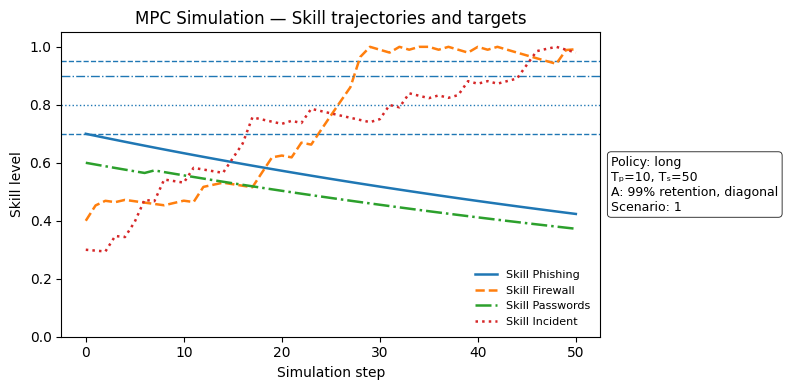}
    \caption{Simulation results for Scenario~1 (mono-competence resources) under the long-term policy.}
    \label{fig:sim_s1_long}
\end{figure}

\begin{figure}[h]
    \centering
    \includegraphics[width=\linewidth]{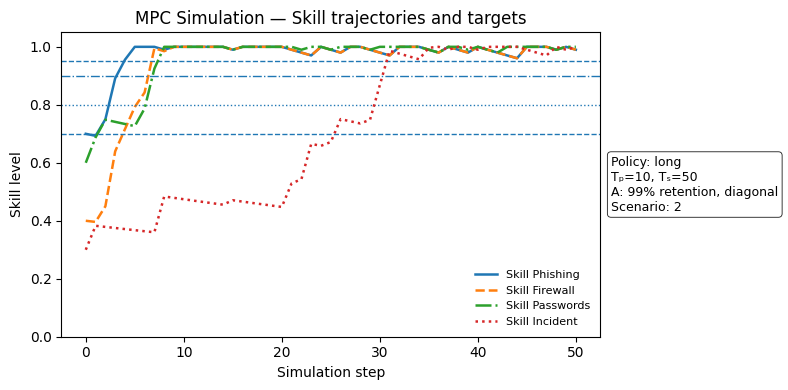}
    \caption{Simulation results for Scenario~2 (semi-clustered competences) under the long-term policy.}
    \label{fig:sim_s2_long}
\end{figure}

\begin{figure}[h]
    \centering
    \includegraphics[width=\linewidth]{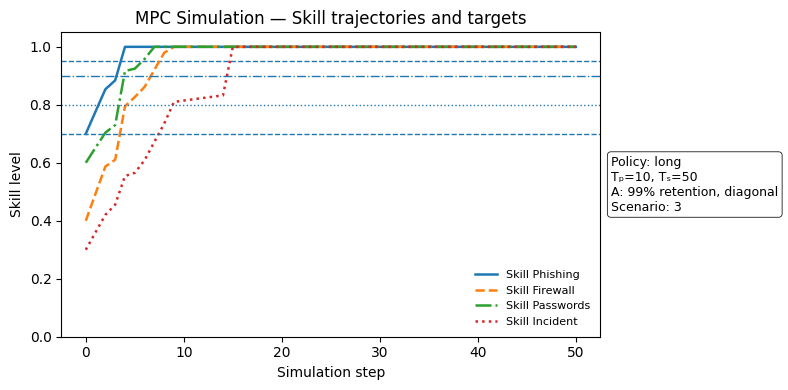}
    \caption{Simulation results for Scenario~3 (multi-competence resources) under the long-term policy.}
    \label{fig:sim_s3_long}
\end{figure}

We chose arbitrary small values to test a MPC optimization. The optimization was resolved every 10 steps, using a Top-$k$ of 8 items generated under a long-term policy. A decay factor of 0.99 was applied to the transition matrix $A$ to model the natural forgetting process, this value illustrates a very small forgetting process and it could be calibrated more wisely. Running the simulation produce the appropriate list of Top-\(k\) and the simulation of learner's choice including a stochastic noise provides some insights. Using the simulation run over 50 time steps and observing the system behavior in Figs.~\ref{fig:sim_s1_long}--\ref{fig:sim_s3_long} illustrate how competences could theoretically progress for a learner.

\begin{figure}[h]
    \centering
    \includegraphics[width=\linewidth]{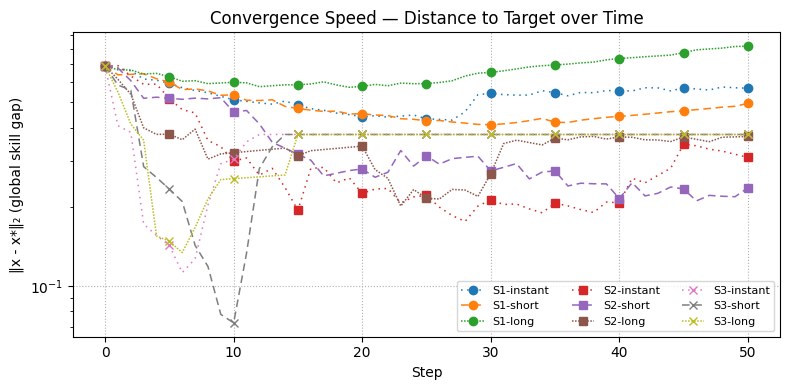}
    \caption{Convergence speed of the TBRS model, expressed as the Euclidean distance over time.}
    \label{fig:conv_speed}
\end{figure}

Of course, these figures are based on synthetic data and cannot be interpreted as representative of real-world experimentation. Rather, they illustrate the type of analytical outcomes that could be expected from such experiments. Likewise, Figure ~\ref{fig:conv_speed} primarily reflects the chosen simulation parameters for the different scenarios, but it also suggests the kind of trends one might observe in a real-world ERS evaluation.

\section{Conclusion}

As recommender systems (RS) continue to shape digital experiences, classical approaches may still exhibit a trajectorial effect, in which user interactions are guided along limited or suboptimal paths. This highlights the need for more adaptive and goal-oriented recommendation strategies. In this context, trajectory-based recommender systems (TBRS) explicitly incorporate notions of user goals and trajectories, aiming to guide users through an item space in a coherent and progressive manner.

We argue that a control-theoretic formulation provides an appropriate and versatile framework for designing such systems. By formalizing TBRS within this framework, a perspective explored in only a few prior studies \parencite{loong_control_2024, chow_control_2024}, we offer a systematic approach that can model user trajectories, accommodate stochastic dynamics, and integrate constraints arising from robustness requirements or data availability. This framework allows the generation of recommendations starting from a minimal set of items, progressively expanding the system as additional resources or usage data become available.

In our educational setting, this approach is particularly relevant: a limited set of initial learning resources can serve as a foundation, while prerequisites, evaluations, and usage traces guide the incremental adaptation of the RS. The flexibility of the control-theoretic approach ensures that evaluations (observations) and initial states can be handled without restrictive assumptions, making it possible to incorporate domain-specific constraints and objectives naturally.

Several important avenues remain for future work. First, empirical validation of TBRS in real-world environments or on offline datasets is needed to assess their effectiveness. Second, the choice of resolution methods within the TBRS framework, balancing robustness, performance, and prediction accuracy, requires further investigation. Third, Top-$k$ recommendation strategies may be oriented toward short-term or long-term objectives, depending on the application context, and exploring the integration of contextual information could substantially enhance performance. Finally, extending the control-theoretic framework to higher-dimensional or more complex recommendation scenarios offers a promising research direction.

Overall, the proposed control-theoretic perspective provides a rigorous, generalizable, and adaptable foundation for TBRS. It not only formalizes the notion of guided user trajectories but also opens the door to innovative, data-driven strategies capable of addressing diverse domains and evolving user needs.

\printbibliography

\end{document}